\documentclass[12pt,a4paper]{article}
\usepackage[T1]{fontenc}
\usepackage[utf8]{inputenc}
\usepackage{lmodern}
\usepackage[margin=2.5cm]{geometry}
\usepackage{amsmath}
\usepackage{amssymb}
\usepackage{booktabs}
\usepackage{array}
\usepackage{tabularx}
\usepackage{graphicx}
\usepackage{hyperref}
\usepackage{url}
\usepackage{parskip}
\usepackage{enumitem}
\usepackage{float}
\usepackage{authblk}

\hypersetup{colorlinks=true,linkcolor=blue,citecolor=blue,urlcolor=blue}

\title{\textbf{A Multi-Modal Dataset for Ground Reaction Force Estimation\\
Using Consumer Wearable Sensors}}

\author[1]{Parvin Ghaffarzadeh\thanks{Corresponding author}}
\author[2]{Debarati Chakraborty}
\author[2]{Koorosh Aslansefat}
\author[3]{Ali Dostan}
\author[2]{Yiannis Papadopoulos}

\affil[1]{Department of Data Science \& Artificial Intelligence and Modelling (DAIM), University of Hull, UK}
\affil[2]{School of Computer Science, University of Hull, UK}
\affil[3]{School of Science \& Technology, Nottingham Trent University, UK}

\date{}

\begin{document}
\maketitle

\begin{abstract}
This Data Descriptor presents a fully open, multi-modal dataset for estimating vertical ground reaction forces (vGRF) from consumer-grade Apple Watch sensors with laboratory force plate ground truth. Ten healthy adults (aged 26--41 years) performed five activities: walking, jogging, running, heel drops and step drops, while wearing two Apple Watch devices positioned at the left wrist and waist. The dataset comprises 492 validated trials with time-aligned inertial measurement unit (IMU) recordings ($\approx$100 Hz) and force plate vGRF (Force\_Z sampled at 1000 Hz). The release includes raw and processed time series, trial-level metadata, quality-control flags, and machine-readable data dictionaries.

Dataset integrity is ensured through trial-level matching manifests linking recordings across modalities using stable identifiers. The primary trial manifest contains 395 triad-complete trials (wrist+waist+force-plate data; 80.3\% of total trials) suitable for comprehensive cross-sensor analyses and reproducible downstream model evaluation. Sensor availability across the full dataset is high: wrist data are available for 461 trials (93.7\%), waist data for 451 trials (91.7\%), and force plate data for 449 trials (91.3\%).

Dataset quality is characterised using a three-phase cross-sensor plausibility and consistency framework (waist$\to$force plate, wrist$\to$waist, wrist$\to$force plate), alongside repeatability metrics for peak vGRF (intraclass correlation coefficient 0.871--0.990) and systematic checks of force ranges and trial completeness. A Monte Carlo sensitivity analysis indicates that correlation-based validation metrics are robust to single-sample timing perturbations at the IMU sampling resolution (mean absolute change correlation $\approx$0.010 for comparisons with $|r| \geq 0.2$ under $\pm$10 ms perturbations, equivalent to $\pm$1 sample at $\sim$100 Hz). All data are available under CC BY 4.0; analysis scripts are archived with the release and mirrored on GitHub. This resource supports reproducible research in wearable biomechanics, benchmarking machine learning vGRF estimation models, and systematic investigation of sensor placement effects using widely available consumer wearables.

\end{abstract}

% ── Background & Summary ──────────────────────────────────────────────────────
\section{Background \& Summary}

Ground reaction forces (GRFs) are widely used to characterise human locomotion, assess musculoskeletal loading, and support applications in sports science, rehabilitation and digital health \cite{winter2009}. Laboratory force plates remain the reference method for GRF measurement, but their size, cost and stationary nature limit widespread use. Wearable inertial sensors offer an accessible alternative, enabling GRF-related signals to be collected during everyday movement. Previous research has demonstrated the feasibility of estimating vertical GRF from wearable inertial sensors during walking and running \cite{jiang2020,wouda2018}. However, benchmark datasets combining consumer wearables with gold-standard force plates remain limited, especially those providing time-aligned recordings at multiple body locations.

Publicly available biomechanical datasets have played a central role in advancing gait analysis, wearable sensor validation, and machine-learning-based estimation of human movement. Early foundational work by Winter \cite{winter1983} established normative datasets for level-ground walking using motion capture and force plates, forming the basis for modern gait research. Subsequent efforts expanded the scope of open-source datasets to include a wider range of locomotor speeds, terrains, and sensor modalities. For example, Moore et al.\ \cite{moore2015} provided perturbation-based walking data, while Fukuchi et al.\ \cite{fukuchi2017,fukuchi2018} released large datasets of walking and running biomechanics across multiple speeds, all captured with motion capture and force plates. Other datasets such as Hu et al.\ \cite{hu2018} and Camargo et al.\ \cite{camargo2021} further broadened the available biomechanical landscape by incorporating ramps, stairs, and transitional gait, often with time-aligned EMG and force-plate measurements to enable multi-modal analyses. Table \ref{tab:comparison} provides a systematic comparison of the present dataset with these and other existing open-source biomechanical datasets.

Recent contributions from \textit{Scientific Data} demonstrate the expanding scope of open source biomechanic datasets. Scherpereel et al.\ \cite{scherpereel2023} provide comprehensive lower-limb data across 31 activities (11 cyclic, 20 non-cyclic) with motion capture, force plates, EMG, and both real and virtual IMU, establishing an important benchmark for multi-sensor biomechanics.

Seong et al.\ \cite{seong2024} presented the MultiSenseBadminton dataset, a multi-modal upper-limb biomechanics dataset designed to assess badminton stroke performance across 25 players. Their dataset combines body tracking (17 IMU sensors including the wrists), eye tracking, motion capture, EMG, and foot pressure sensors to evaluate movement quality and skill-level differences. Although sport-specific, this work demonstrates the feasibility of time-aligning multiple consumer-grade and research-grade sensors during fast, high-acceleration activities and highlights the potential of wrist-worn sensors for analysing dynamic upper-limb movement, providing methodological insights relevant to wrist-based sensing in general.

Recent large-scale data aggregation efforts have further expanded the biomechanics dataset landscape. The AddBiomechanics project \cite{werling2024} provides a massive repository of motion capture and force plate data compiled from multiple laboratories, enabling musculoskeletal modeling at scale. However, AddBiomechanics does not include wearable sensor recordings, limiting its direct applicability to consumer device validation. Similarly, the UNDERPRESSURE dataset \cite{fang2022} combines motion capture with plantar pressure insoles, offering insights into foot-ground interaction dynamics, but lacks time-aligned force plate ground truth. These complementary resources highlight the diversity of measurement modalities in biomechanics research, while underscoring the specific gap that consumer wearable-force plate paired datasets address.

While these datasets collectively offer rich resources for gait analysis, neuromechanics, and movement classification, none of them focus on wrist-worn consumer devices for estimating lower-limb ground reaction forces. Existing public datasets primarily rely on proximal sensors (e.g., shank, thigh, pelvis) or full-body motion capture, while wrist placement remains underrepresented despite its ubiquity in consumer health technology.

This dataset addresses these gaps by combining wrist-worn consumer IMUs with laboratory grade force plate vGRF across walking, jogging, running, and impact-based tasks (heel drops and step drops). Dual placement (wrist and waist) enables systematic investigation of sensor placement effects and supports research into distal (wrist-only) vGRF inference.

\begin{table}[H]
\centering
\caption{Comparison of the present dataset with existing open-source biomechanical datasets combining wearable sensors and ground truth measurements.}
\label{tab:comparison}
\resizebox{\textwidth}{!}{%
\begin{tabular}{lllllll}
\toprule
\textbf{Dataset} & \textbf{Year} & \textbf{Sensor Placement} & \textbf{Force Plate} & \textbf{Primary Activities} & \textbf{N} & \textbf{Public} \\
\midrule
Winter \cite{winter1983} & 1983 & Motion capture & Yes & Walking & 1 & Yes \\
Moore et al.\ \cite{moore2015} & 2015 & Motion capture & Yes & Walking, perturbations & 15 & Yes \\
Fukuchi et al.\ \cite{fukuchi2017} & 2017 & Lower-limb markers & Yes & Running (multiple speeds) & 28 & Yes \\
Fukuchi et al.\ \cite{fukuchi2018} & 2018 & Lower-limb markers & Yes & Walking (multiple speeds) & 42 & Yes \\
Hu et al.\ \cite{hu2018} & 2018 & Shank, thigh & Yes & Stairs, ramps, level ground & 10 & Yes \\
Schreiber \& Moissenet \cite{schreiber2019} & 2019 & Lower-limb markers, EMG & Yes & Walking (multiple speeds) & 50 & Yes \\
Lencioni et al.\ \cite{lencioni2019} & 2019 & Lower-limb markers & Yes & Walking, stairs & 50 & Yes \\
Camargo et al.\ \cite{camargo2021} & 2021 & Lower-limb markers & Yes & Stairs, ramps, transitions & 22 & Yes \\
Reznick et al.\ \cite{reznick2021} & 2021 & Lower-limb markers & Yes & Variable-speed locomotion & 10 & Yes \\
Scherpereel et al.\ \cite{scherpereel2023} & 2023 & Shank, thigh, pelvis, foot & Yes & 11 cyclic + 20 non-cyclic & 12 & Yes \\
Seong et al.\ \cite{seong2024} & 2024 & Body tracking IMUs (17 sensors), foot pressure & No & Badminton strokes & 25 & Yes \\
Ghaffarzadeh et al.\ (present work) \cite{ghaffarzadeh2025} & 2025 & Left wrist, waist & Yes & Walk, jog, run, drops & 10 & Yes \\
\bottomrule
\end{tabular}%
}
\end{table}

% ── Methods ───────────────────────────────────────────────────────────────────
\section{Methods}

\subsection{Ethics Approval and Consent}

This study received ethical approval from the University of Hull Faculty of Health Sciences Research Ethics Committee (reference: FHS-24-25.036). All participants provided written informed consent prior to participation, including explicit consent for public data sharing in fully de-identified form. All methods were performed in accordance with relevant guidelines and regulations.

\subsubsection*{Recruitment, Screening, and Participant Privacy}

Interested individuals accessed an online expression-of-interest form and provided basic, non-sensitive information to assess eligibility (age 18--65 years; no recent musculoskeletal injury; no bone/musculoskeletal/neurological condition affecting safe performance; and no cardiovascular contraindications to moderate-intensity activity). Eligible individuals were contacted by the student investigator (P.G.) via email, provided with the Participant Information Sheet, and given at least 24 hours to review the study and ask questions before providing written consent in person at the Sport Science laboratory prior to any data collection. A pre-exercise medical questionnaire was completed prior to testing, and participants could withdraw at any time without providing a reason.

To protect participant privacy, all shared data are de-identified in accordance with the UK GDPR and the UK Data Protection Act 2018. No direct identifiers (e.g., names, addresses, email/phone contact details, or full dates of birth) are included in the released dataset. Participants are identified only by pseudonymous codes (e.g., P1--P10). Anthropometric variables (e.g., age, height, weight, BMI) are provided in the dataset metadata tables; in the current public archive, basic anthropometrics (birth year, height, body mass) also appear in participant folder names for navigation, while no direct identifiers (names, contact details, full dates of birth) are included. The linkage between participant codes and identities (including consent and contact information) is stored separately from the research data on access-restricted, password-protected University systems and is accessible only to the research team.

\subsection{Participants}

Ten healthy adults (6 male, 4 female) participated in the study (Table \ref{tab:demographics}). Participants were recreationally active ($\geq$ 2 hours/week of moderate-intensity exercise) with no self-reported lower-limb injury in the 6 months prior to participation that would impair movement, and no self-reported diagnosis of bone, musculoskeletal, or neurological conditions affecting gait or balance. Inclusion criteria allowed participants aged 18--65 years (actual sample: 26--41), ability to walk and perform moderate-intensity activities without pain, and no conditions that would preclude safe participation. Footwear was not standardised; participants wore their own shoes, and shoe size and shoe type were recorded for all participants (Table \ref{tab:demographics} and accompanying metadata). ``Trainer'' denotes athletic running or fitness shoes with a compliant foam midsole and flexible upper designed for physical activity; ``Casual'' denotes everyday shoes not designed for running, typically featuring a leather or synthetic upper with a stiffer midsole and less cushioning (e.g., a leather lace-up walking shoe). Only participant P7 wore casual footwear; all other participants wore athletic trainers.

\begin{table}[H]
\centering
\caption{Participant demographics ($n=10$).}
\label{tab:demographics}
\resizebox{\textwidth}{!}{%
\begin{tabular}{llllllll}
\toprule
\textbf{Participant} & \textbf{Sex} & \textbf{Age (yr)} & \textbf{Height (cm)} & \textbf{Weight (kg)} & \textbf{BMI (kg/m$^{2}$)} & \textbf{Shoe size} & \textbf{Shoe type} \\
\midrule
P1  & M & 35 & 186.5 & 95.8  & 27.5 & 44   & Trainer \\
P2  & F & 26 & 165.5 & 74.5  & 27.2 & 40   & Trainer \\
P3  & M & 36 & 172.0 & 67.8  & 22.9 & 41   & Trainer \\
P4  & F & 32 & 176.0 & 68.6  & 22.1 & 39   & Trainer \\
P5  & M & 39 & 174.0 & 74.0  & 24.4 & 43   & Trainer \\
P6  & F & 34 & 162.0 & 68.2  & 26.0 & 37   & Trainer \\
P7  & M & 36 & 185.5 & 85.2  & 24.8 & 41   & Casual  \\
P8  & M & 31 & 175.0 & 68.8  & 22.5 & 43.5 & Trainer \\
P9  & F & 33 & 167.0 & 60.2  & 21.6 & 37   & Trainer \\
P10 & M & 41 & 173.0 & 76.4  & 25.5 & 42   & Trainer \\
\midrule
Mean $\pm$ SD & --- & $34.3\pm4.2$ & $173.7\pm7.9$ & $74.0\pm10.1$ & $24.5\pm2.1$ & --- & --- \\
Range         & --- & 26--41 & 162.0--186.5 & 60.2--95.8 & 21.6--27.5 & --- & --- \\
\bottomrule
\end{tabular}%
}
\end{table}

\subsection{Data Collection Protocol}

Each participant completed a single laboratory visit (one day). The session lasted approximately 60 minutes including setup, familiarisation, and data collection. To quantify within-session repeatability, each activity was repeated multiple times depending on the participant's ability (typically 5--12 trials per activity, occasionally up to 16).

\subsubsection*{Experimental Setup}

Participants wore two Apple Watch devices (Series 5 or later): one on the dominant wrist (standard watch position, typically the left wrist) and one secured at the waist using an elastic belt positioned at the anterior waist, inferior to the navel, close to the body's centre of mass (COM). Watch firmware and iOS versions were standardised across sessions (iOS 16+ with watchOS 9+). Ground reaction forces were recorded using an AMTI OR6-7 force plate (1000 Hz sampling rate, sensitivity 10 N), mounted flush with the laboratory floor. Participants' height and weight were measured at the beginning of the first session using a calibrated stadiometer and digital scale. Footwear was not standardised; shoe type and shoe size were recorded for all participants and are provided in the supplementary metadata.

\paragraph{Sensor placement, securing method, and wrist side.}
Two Apple Watch devices (Series 5 or later) were used concurrently: one at the wrist and one at the waist. Watches were placed by the researcher (P.G.) to standardise position and tightness across participants. The wrist device was worn on the left wrist in a standard watch position (dorsal wrist, proximal to the ulnar styloid). The watch band was tightened to a snug, non-slipping fit, minimising relative motion between device and skin without restricting circulation. The waist device was secured using an elastic belt worn over clothing and tightened to prevent bounce. The device was oriented with the watch face outward and positioned inferior to the navel on the anterior trunk, close to the body's centre of mass (COM). The figure schematic is illustrative; the anterior waist position inferior to the navel is the operative reference for all analyses and metadata. Before each trial, the investigator confirmed the pre-assigned activity label and trial index with the participant to ensure consistency with the predefined trial order defined in the acquisition protocol. Force-plate acquisition was initiated by the investigator immediately prior to movement onset, while wearable recordings were started on both Apple Watches at the beginning of the same trial and stopped immediately after completion. Each wearable file stores a unique UUID together with millisecond-resolution timestamps for provenance, and trials are linked across modalities using participant ID, activity label, and the paired IMU identifier with the corresponding force-plate trial number, as recorded in the trial manifests.

\paragraph{Standing reference segment.}
At the beginning of each session, participants stood still for approximately 5 s to verify sensor streaming stability. These segments are included in the raw IMU recordings as part of the pre-activity buffer (no separate calibration file is required; the standing segment is the first portion of each trial file).

\paragraph{Reporting considerations for wearable inertial sensing.}
We report sensor placement, securing method, coordinate conventions, sampling characteristics, and preprocessing choices following the International Society of Biomechanics (ISB) recommendations for wearable inertial measurement--based human motion analysis, with a focus on transparency and reproducibility (Cereatti et al., 2024).

\subsubsection*{Sensor Specification and Data Acquisition}

Apple Watch devices provide tri-axial acceleration and angular velocity through the Core Motion framework. The accelerometer has a nominal measurement range of $\pm$8 g, and the gyroscope operates within $\pm$2000$^\circ$/s. Sensor data were sampled at approximately 100 Hz and recorded using a custom-developed iOS/watchOS application (Swift 5.8+), which stored timestamped sensor readings in CSV format with millisecond precision. Each recording included the full trial duration, with short pre- and post-activity buffer periods to ensure complete capture of the movement.

The Apple Watch coordinate system follows the standard device reference frame: $+X$ (lateral, toward the screen edge), $+Y$ (longitudinal, toward the digital crown), and $+Z$ (perpendicular to the device surface). The recorded \textbf{userAcceleration} signals represent gravity-compensated linear acceleration estimated by Apple's internal sensor-fusion algorithm, which combines accelerometer and gyroscope data to estimate device orientation and remove the gravity component. Angular velocity was recorded using the gyroscope rotationRate signals. These signals were recorded simultaneously for both the wrist- and waist-mounted devices throughout each trial.

Ground reaction force data were recorded simultaneously using an AMTI OR6-7 force plate connected to the NetForce acquisition system at a sampling frequency of 1000 Hz. Force-plate data were exported as CSV files containing the full AMTI outputs (forces, moments, and centre-of-pressure). In this work, we use and report the vertical component (Force\_Z) as vGRF for time-alignment, quality control, and validation.

\paragraph{Task descriptions and trial acceptance (overground protocol).}
Participants completed overground trials along a straight 10 m walkway. Each locomotor trial consisted of a single pass at a self-selected speed, with speed recorded using timing gates (see Section ``Self-selected speed''). Participants were instructed to look forward, maintain their natural gait, and avoid visibly targeting the force plate.

Trials were accepted only when a clean foot contact occurred on the force plate. If a foot did not land cleanly on the plate (e.g., partial contact, visible targeting, or double contact), the trial was rejected and immediately repeated. A brief familiarisation period was provided for each activity before recorded trials commenced.

A single AMTI OR6-7 force plate was used, and participants were instructed to contact the plate with their \textbf{right foot} during locomotor passes.

\paragraph{Walking.}
Participants walked at a comfortable, self-selected speed along a 10 m overground walkway.

\paragraph{Jogging.}
Participants jogged at a comfortable easy-run intensity, faster than walking but below running pace.

\paragraph{Running.}
Participants ran at a self-selected pace, faster than jogging but below sprinting intensity.

\paragraph{Heel drop (impact task).}
Participants stood on the force plate, rose onto the forefoot (plantarflexion), then dropped their heels to contact the plate, producing a short-duration impact transient. Participants were instructed to keep the upper body upright with arms crossed over the chest, and to reset between repetitions.

\paragraph{Step drop (impact task).}
Participants stepped down from a 20 cm step platform placed adjacent to a single AMTI OR6-7 force plate, landing naturally (dominant foot first), then returned to the step for the next repetition. Participants were instructed to avoid jumping and to keep the motion controlled but natural. Trials were rejected and immediately repeated if the foot did not land cleanly on the force plate. A brief familiarisation period was provided before recorded trials commenced.

\subsection{Data Processing}

Force-plate signals were low-pass filtered using a 4th-order Butterworth filter (20 Hz cutoff). IMU accelerometer and gyroscope signals were filtered using a 2nd-order Butterworth filter (10 Hz cutoff). These cutoffs were selected to retain the dominant biomechanically relevant content for walking/running and impact tasks while reducing sensor noise. Filtering was applied for correlation-based quality control, validation analyses, and visualisation; all raw (unfiltered) IMU and force-plate signals are also provided to support analyses of higher-frequency impact transients or user-defined preprocessing. For impact-focused analyses (e.g., heel drops and step drops), users should preferentially use the unfiltered IMU signals, as higher-frequency components may be attenuated by the 10 Hz filter. To confirm that filtering did not substantially distort key impact features, we compared peak force magnitudes and event timing between filtered and unfiltered signals for a representative subset of trials ($n = 50$). Peak force differences were $<$2.5\%, and event timing differences were $<$5 ms. Acceleration magnitude was computed as $a_{\mathrm{mag}} = \sqrt{a_x^2 + a_y^2 + a_z^2}$ to reduce orientation dependence for cross-sensor comparisons. \\
Stance phase was identified from the vertical GRF as the longest contiguous segment where Force\_Z exceeded 50 N, with a minimum duration of 150 ms for cyclic locomotor activities, following the widely adopted criterion of Hreljac and Marshall \cite{hreljac2000}. For impact tasks (heel drops, step drops), a shorter minimum duration threshold (50 ms) was used to accommodate brief contact periods, consistent with typical contact durations reported in drop-landing studies \cite{mcnitt2001}. Heel-tap marker detection for temporal alignment is described in the next section. Trial identifiers were embedded during acquisition using unique UUIDs to enable one-to-one matching of wrist IMU, waist IMU, and force-plate recordings; metadata include participant characteristics, device placement, activity labels, and quality-control flags.

\subsection{Quality Control}

Quality control included: (i) stance detection verification; (ii) physiological plausibility checks of peak forces \cite{winter2009,cavanagh1980}; (iii) detection of missing data or signal dropouts; (iv) time-alignment completeness checks; and (v) screening for accelerometer near-range behavior and potential clipping. Trials with incomplete sensor recordings, failed time-alignment, or implausible force values were excluded from the final dataset. A systematic accelerometer saturation analysis and per-activity summary are provided in Usage Notes (Accelerometer saturation). The final released dataset contains 492 validated trials across all sensor modalities.

\subsubsection*{Trial Matching and Manifests}

We use the term trial manifest to refer to a machine-readable index table (CSV) that maps each trial's files across modalities (wrist IMU, waist IMU, force plate), including file paths, trial identifiers, availability flags, and QC metadata. The dataset includes trial-level matching manifests enabling identification of corresponding force plate, waist IMU, and wrist IMU recordings for each trial. Each trial is assigned a unique identifier composed of participant code, activity type, and trial number (e.g., P1\_Walk\_ID5). Manifests are provided as CSV files with columns for trial identifier, file paths, availability flags (wrist/waist/force-plate), and quality-control metadata. The primary manifest (trial\_manifest.csv) contains 492 validated trials, of which 395 (80.3\%) are triad-complete (all three modalities present). Two supplementary manifests provide trial count summaries: trial\_counts\_all.csv (all 492 trials) and trial\_counts\_triad\_complete.csv (395 triad-complete trials).

\subsubsection*{Temporal Alignment}

Temporal alignment between wearables and force plate used a hybrid strategy combining event-based detection (when a distinct marker was present) with stance-feature alignment and cross-correlation validation. The alignment pipeline was executed on all matched force+waist+wrist recordings present in the raw collection (678 matched triplets), producing 598 successfully aligned outputs (88.2\% success). The released 492 validated trials reflect additional curation for sensor availability, stance validation, plausibility checks, and metadata completeness; therefore, counts differ by design. We verified that manuscript counts (678$\to$598 aligned outputs; 492 validated; 395 triad-complete) match the released trial manifests.

\paragraph{Heel-tap marker (optional, event-based).}
At the start of some trials, a brief ground contact marker was performed to provide an identifiable alignment cue across modalities. Specifically, the participant performed a short, deliberate foot tap on the force plate while standing still. This produces a sharp vertical force impulse and a whole-body mechanical transient that propagates through the lower limbs and trunk, detectable at the waist and often also at the wrist due to impact transmission through the upper limb segments. When present, we detect this cue within the first 2 s and use it as an initial alignment anchor; when absent or ambiguous, we rely on stance-feature alignment (below).

\paragraph{Heel-tap marker detection (event-based).}
When a heel-tap marker was present, we searched the first 2 s of each recording for a distinct impulse observable in both modalities. The force-plate marker time was defined as the first sustained threshold crossing in vertical force (Force\_Z $>$ 100 N for $\geq$ 20 ms), followed by a prominent local maximum. A marker was accepted as ``clear'' only if a co-occurring peak in IMU acceleration magnitude was present within a bounded temporal window around the force-plate impulse.

\paragraph{Locomotor stance-feature alignment (walk/jog/run):}
For walking, jogging, running candidate alignment windows were derived from stance using the longest contiguous segment where Force\_Z exceeded a contact threshold (80 N) with minimum duration 150 ms. Within this window, we constructed force and IMU proxy signals based on gradients of Force\_Z and acceleration magnitude and estimated lag via cross-correlation. The final lag was applied to produce aligned time series at 100 Hz.

The contact threshold of 80 N for locomotor alignment window selection was chosen to reliably exclude near-zero baseline noise while remaining below typical toe-off force levels in walking and running, consistent with thresholds reported by Cavanagh and Lafortune \cite{cavanagh1980} and Weyand et al.\ \cite{weyand2010}. The 150 ms minimum duration criterion follows the widely adopted stance detection convention of Hreljac and Marshall \cite{hreljac2000}.

\paragraph{Proxy signal construction for alignment.}
Within the detected contact window, we construct proxy signals that emphasise impact and loading transitions: for the force plate we use the first-order temporal gradient (discrete derivative) of Force\_Z after filtering; for the IMU we use the gradient of acceleration magnitude. These gradient signals accentuate rapid transitions (initial contact, push-off) that are common to both modalities, yielding sharper cross-correlation peaks and more robust lag estimates than using the raw signals directly. We then compute the lag that maximises the normalised cross-correlation between these proxy signals over a bounded lag range ($\pm$150 ms) and apply this lag to produce aligned time series at 100 Hz. The per-trial alignment log records the chosen lag, method used, overlap coverage, and cross-correlation quality.

\paragraph{Impact tasks (heel drops/step drops).}
For heel-drop and step-drop trials, contact durations are shorter; therefore, we used a reduced minimum contact duration (50 ms) and a relaxed contact threshold (40 N) for stance-window detection. The same proxy construction and cross-correlation validation were applied within the detected contact window.

For impact tasks, the reduced contact threshold of 40 N and minimum duration of 50 ms reflect the shorter, more transient contact profiles typical of drop landings, in line with recommendations for detecting impact transients in drop-landing studies \cite{mcnitt2001,dempsey2009}. The heel-tap detection threshold of 100 N was selected to identify the deliberate impact event reliably while excluding baseline noise, following the convention of using approximately 10--20\% of expected peak GRF for impact event detection \cite{dempsey2009,pataky2008}.

\paragraph{What the pipeline guarantees.}
The alignment procedure outputs time-aligned signals and a per-trial alignment log reporting estimated lag, method selection, overlap coverage, and correlation quality metrics. Monte Carlo time-alignment sensitivity analysis quantifies how bounded single-sample lag perturbations ($\pm$10 ms at 100 Hz) propagate through the three-phase correlation framework; this characterises metric sensitivity rather than absolute timing accuracy.

\paragraph{Synchronisation scope and limitations.}
Force plate acquisition was initiated by the researcher immediately prior to movement onset. Upon a verbal cue from the researcher, the participant sequentially pressed the Start button on each Apple Watch device (wrist and waist) using the MyCoreMotion app, introducing small inter-device delays at acquisition onset. The Apple Watch and force plate do not share a hardware clock, and these sequential manual triggers introduce offsets that are subsequently estimated and corrected by the post-hoc alignment pipeline. Our approach therefore provides post-hoc temporal alignment using event cues (when present) and stance-feature matching with cross-correlation verification. Such post-hoc alignment methods are commonly used in multi-sensor workflows where a shared hardware clock is unavailable and have been adopted in comparable behavioural datasets (e.g., the Gaze-in-Wild dataset \cite{kothari2020}). This procedure does not constitute validation against an independent time-synchronised reference system; accordingly, alignment outputs and QC metrics should be interpreted as supporting internal consistency and usability for downstream modelling rather than as proof of absolute timing accuracy. The per-trial alignment log records the estimated lag, alignment method, overlap coverage, and cross-correlation quality, allowing users to re-align signals using alternative methods if required.

% ── Data Records ──────────────────────────────────────────────────────────────
\section{Data Records}

The dataset is publicly available on Zenodo as a versioned release (version DOI: 10.5281/zenodo.17376717; concept DOI: 10.5281/zenodo.17376716). The dataset should be cited as: Ghaffarzadeh et al.\ (Zenodo) \cite{ghaffarzadeh2025}.

\subsection{File Organisation}

Data are organised by participant (P1--P10). Each participant folder contains subfolders for force plate recordings (Activity Type/), wrist IMU recordings (Wrist/), and waist IMU recordings (Waist/). Trials are stored as a separate CSV file with standardised naming that includes activity, sensor location, trial index, and a unique UUID to enable cross-sensor matching:

\begin{center}
[Activity]\_[Sensor]\_Trial\_[UUID]\_ID\_[Number]\_[Timestamp].csv.
\end{center}

\begin{figure}[H]
  \centering
  \includegraphics[width=0.75\textwidth]{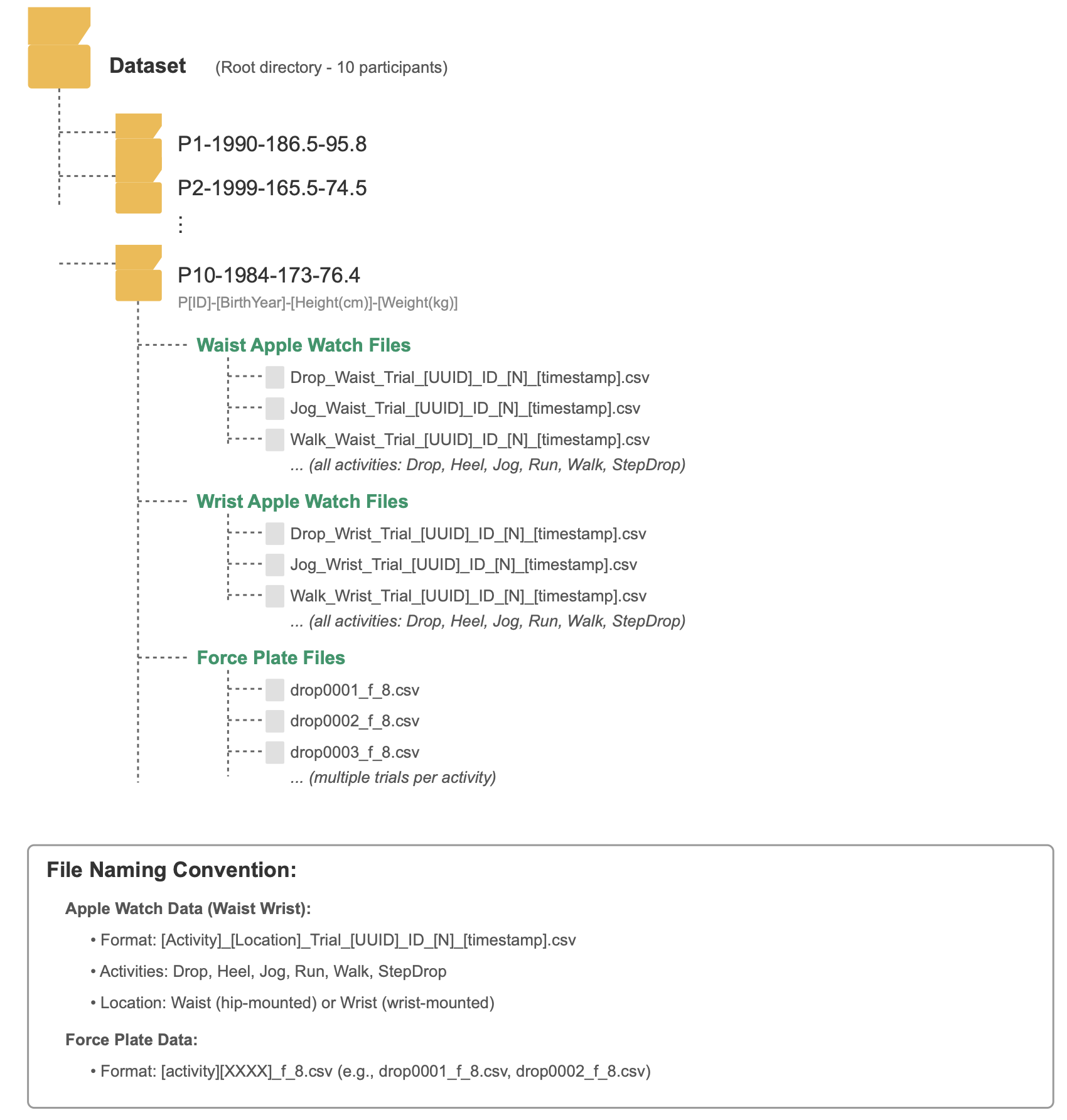}
  \caption{Dataset folder structure and organisation. Hierarchical organisation showing 10 participant subfolders (P1--P10), each containing time-aligned IMU sensor files for wrist and waist positions across five activities. Each trial is identified by a unique UUID enabling precise cross-sensor matching. The alignment pipeline produced 598 aligned outputs from 678 matched triplets (pre-curation). The released dataset comprises 492 validated trials documented in trial\_manifest.csv (395 triad-complete).}
  \label{fig:folder}
\end{figure}

Participant folders follow the convention P[ID]-[BirthYear]-[Height]-[Weight] to support convenient access to anthropometrics during analysis. Only year of birth is included (not full date of birth). The released dataset contains no names, addresses, contact details, or other direct identifiers; participant linkage information is stored separately under restricted access within the research team. Figure \ref{fig:folder} illustrates the dataset folder hierarchy and structure.

\paragraph{Privacy note.}
Participant folders in the public archive currently follow the naming convention P[ID]-[BirthYear]-[Height]-[Weight] to facilitate rapid manual navigation and linkage with participant metadata tables. These attributes represent general demographic variables and do not include direct identifiers such as names, addresses, contact details, or full dates of birth. All participant records are pseudonymised using participant codes (P1--P10) in accordance with the UK GDPR and institutional data-protection policies. If additional anonymisation is preferred by downstream users, the dataset can readily be distributed in an alternative archive structure using pseudonymous participant IDs only (P01--P10), with demographic variables provided exclusively in the metadata tables.

\subsection{Data Files}

\begin{itemize}
  \item \textbf{Raw time series:} Force plate (1000 Hz) and IMU (100 Hz) recordings in CSV format with timestamps, 3-axis acceleration, and 3-axis angular velocity (IMU). Force plate files contain the full AMTI export, including three-axis forces (Force\_X, Force\_Y, Force\_Z (vGRF)), moments, and centre-of-pressure coordinates.
  \item \textbf{Processed time series:} Filtered signals (20 Hz force, 10 Hz IMU) with derived features (acceleration magnitude and stance phase indicators). Analyses and alignment procedures primarily use the vertical ground reaction force component (Force\_Z).
  \item \textbf{Trial manifests:} trial\_manifest.csv (492 trials), trial\_counts\_all.csv, \\trial\_counts\_triad\_complete.csv (395 triad-complete).
  \item \textbf{Metadata:} Participant demographics (participant\_demographics.csv), per-trial quality flags, alignment logs, and data dictionaries.
\end{itemize}

\subsection{Data Dictionary}

Each CSV file follows a standardised column schema documented in data\_dictionary.json, including variable names, units, descriptions, and valid ranges. Key variables include:

\begin{itemize}
  \item \textbf{timestamp:} Millisecond-precision time (Apple Watch) or time in seconds (force plate)
  \item \textbf{userAccelerationX/Y/Z:} Body acceleration (g) with gravity removed. This is an API-derived signal from Apple's Core Motion sensor-fusion pipeline, not raw specific force; gravity compensation is performed internally by the device, and the compensation method is not user-controllable. Orientation-independent acceleration magnitude is provided as a separate derived variable.
  \item \textbf{rotationRateX/Y/Z:} Angular velocity (rad/s)
  \item \textbf{Force\_Z:} Vertical ground reaction force (N)
  \item \textbf{acceleration\_magnitude:} Euclidean norm of 3-axis acceleration
  \item \textbf{stance\_flag:} Binary indicator (1=stance phase, 0=otherwise)
\end{itemize}

\subsubsection*{File Format Details}

\paragraph{IMU Files.}
CSV format with header row. Key columns include timestamp, tri-axial acceleration (gravity-corrected), tri-axial gyroscope, acceleration magnitude, activity label, trial UUID, participant ID, and sensor placement. Sampling rate approximately 100 Hz.

\paragraph{Force Plate Files.}
CSV format with metadata rows 1--26, header row 27, and data beginning at row 28. Sampling rate 1000 Hz. Key columns include time, three-axis forces (Force\_X, Force\_Y, Force\_Z), three-axis moments, and centre of pressure coordinates. Although the full AMTI force-plate channels are provided, the present dataset description and validation analyses focus primarily on the vertical ground reaction force component (Force\_Z), which is commonly used in wearable GRF estimation studies.

\paragraph{Data Dictionary.}
Machine-readable CSV and JSON files documenting all column names, descriptions, units, sampling rates, valid ranges, and coordinate system conventions.

\subsection{Trial Counts and Availability}

The dataset contains 492 validated trials distributed across 10 participants and 5 activities (Table \ref{tab:trial_counts_all}). Per-participant trial counts range from 31 to 56 (mean 49.2 $\pm$ 6.7 SD). Of the 492 total trials, 395 (80.3\%) are triad-complete with all three modalities available. Table \ref{tab:trial_counts_triad} provides the triad-complete activity-specific breakdown.

\begin{table}[H]
\centering
\caption{Trial counts per participant and activity (all 492 validated trials).}
\label{tab:trial_counts_all}
\begin{tabular}{lcccccc}
\toprule
\textbf{Participant} & \textbf{Walk} & \textbf{Jog} & \textbf{Run} & \textbf{Heel Drop} & \textbf{Step Drop} & \textbf{Total} \\
\midrule
P1  & 10 & 10 & 7  & 11 & 10 & 48  \\
P2  & 11 & 10 & 10 & 10 & 10 & 51  \\
P3  & 9  & 10 & 10 & 10 & 10 & 49  \\
P4  & 11 & 10 & 10 & 11 & 10 & 52  \\
P5  & 11 & 10 & 11 & 11 & 11 & 54  \\
P6  & 11 & 10 & 10 & 10 & 10 & 51  \\
P7  & 10 & 11 & 10 & 10 & 11 & 52  \\
P8  & 5  & 12 & 10 & 10 & 11 & 48  \\
P9  & 6  & 6  & 7  & 7  & 5  & 31  \\
P10 & 12 & 11 & 12 & 10 & 11 & 56  \\
\midrule
\textbf{TOTAL} & \textbf{96} & \textbf{100} & \textbf{97} & \textbf{100} & \textbf{99} & \textbf{492} \\
\bottomrule
\end{tabular}
\end{table}

\begin{table}[H]
\centering
\caption{Trial counts per participant and activity (395 triad-complete trials only).}
\label{tab:trial_counts_triad}
\begin{tabular}{lcccccc}
\toprule
\textbf{Participant} & \textbf{Walk} & \textbf{Jog} & \textbf{Run} & \textbf{Heel Drop} & \textbf{Step Drop} & \textbf{Total} \\
\midrule
P1  & 7  & 0  & 6  & 11 & 10 & 34  \\
P2  & 9  & 10 & 8  & 7  & 9  & 43  \\
P3  & 9  & 10 & 10 & 10 & 10 & 49  \\
P4  & 11 & 10 & 10 & 11 & 10 & 52  \\
P5  & 11 & 10 & 11 & 0  & 0  & 32  \\
P6  & 9  & 7  & 9  & 8  & 9  & 42  \\
P7  & 10 & 7  & 10 & 9  & 10 & 46  \\
P8  & 5  & 0  & 2  & 10 & 11 & 28  \\
P9  & 6  & 6  & 6  & 4  & 5  & 27  \\
P10 & 8  & 11 & 7  & 9  & 7  & 42  \\
\midrule
\textbf{TOTAL} & \textbf{85} & \textbf{71} & \textbf{79} & \textbf{79} & \textbf{81} & \textbf{395} \\
\bottomrule
\end{tabular}
\end{table}

\begin{figure}[H]
  \centering
  \includegraphics[width=\textwidth]{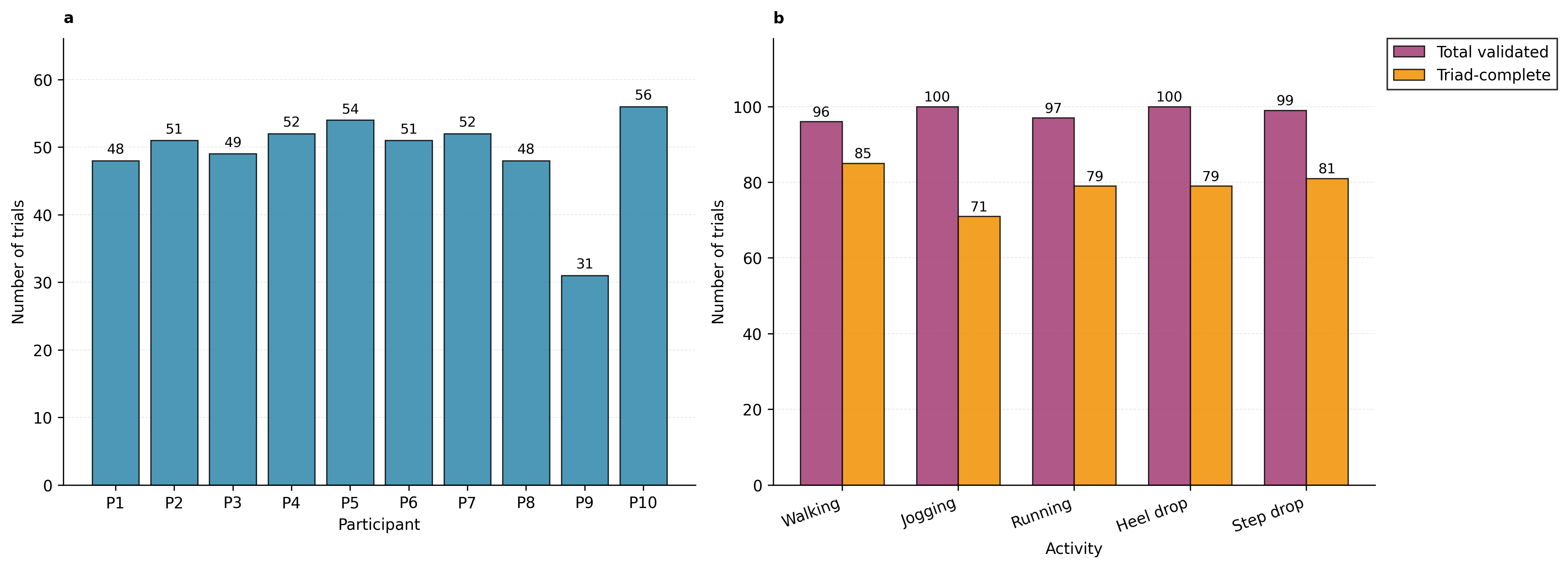}
  \caption{Trial distribution overview. (A) Per-participant trial counts for all 492 validated trials. (B) Activity-specific trial counts showing triad-complete (all sensors) versus total validated trials. The dataset contains 395 triad-complete trials (80.3\%) suitable for comprehensive cross-sensor analyses.}
  \label{fig:trial_dist}
\end{figure}

Figure \ref{fig:trial_dist} provides a visual summary of trial distribution.

\subsection{Data Overview}

To provide a high-level characterisation of the dataset, Figure 3 shows the experimental setup and representative time-aligned IMU and force plate signals for each activity. Panel (A) illustrates the dual-sensor configuration with the waist-worn Apple Watch secured on an elastic belt and the wrist-worn device on the left wrist. The right panels demonstrate activity-specific biomechanical signatures: walking exhibits a characteristic double-peak force pattern, running a single higher-amplitude peak, and drop tasks a rapid high-force transient (2.5--4.0$\times$ body weight). Temporal alignment across all three modalities is evident from the co-occurring peaks in force plate and acceleration signals.

\subsubsection*{Technical Validation}

Technical validation focused on confirming the consistency and quality of the recorded signals.

\subsubsection*{Peak vGRF Repeatability}

% Figure 3: no caption in Word document -- placed inline before Peak vGRF Repeatability paragraph
\begin{figure}[H]
  \centering
  \includegraphics[width=\textwidth]{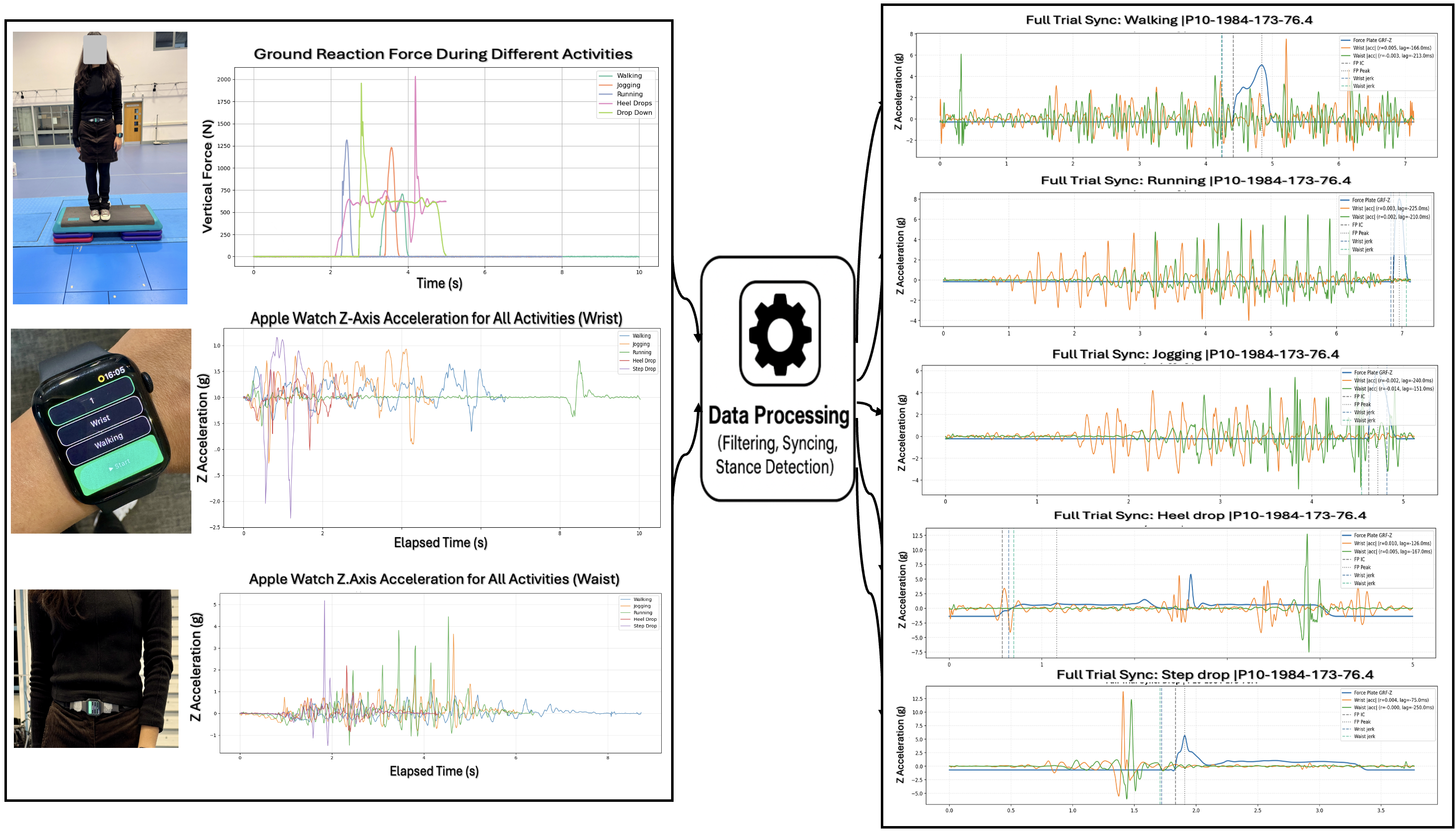}
  \caption{Figure 3. Data collection setup and representative time-aligned signals. Left: (A) Experimental setup showing the waist-worn Apple Watch secured with an elastic belt at the anterior waist (below the navel) and the wrist-worn device on the left wrist; (B) participant standing on the 20 cm step platform used for step-drop trials; (C) MyCoreMotion data collection interface; (D) processing pipeline overview. Right: Representative time-aligned signals showing vertical ground reaction force from the force plate (Force\_Z, black), waist acceleration (blue), and wrist acceleration (red). Vertical lines indicate initial contact ($>$50 N). Walking shows a double-peak pattern, running a single higher-amplitude peak, and drop tasks a rapid high-force onset ($\sim$2.5--4.0$\times$ body weight).}
  \label{fig:overview}
\end{figure}

To assess measurement repeatability, we computed intraclass correlation coefficients (ICC, two-way mixed effects, absolute agreement) for peak vGRF across repeated trials within each activity \cite{hopkins2000,koo2016}. ICC values ranged from 0.871 (heel drops) to 0.990 (running), indicating excellent repeatability (Table \ref{tab:repeatability}). Coefficients of variation were below 8\% for all activities, confirming consistent force plate measurements across sessions.

\begin{table}[H]
\centering
\caption{Repeatability metrics for peak vGRF.}
\label{tab:repeatability}
\begin{tabular}{llll}
\toprule
\textbf{Activity} & \textbf{ICC (95\% CI)} & \textbf{CV (\%)} & \textbf{N} \\
\midrule
Walking   & 0.962 (0.941--0.976) & 4.2 & 96  \\
Jogging   & 0.981 (0.972--0.987) & 3.1 & 100 \\
Running   & 0.990 (0.985--0.993) & 2.3 & 97  \\
Heel Drop & 0.871 (0.823--0.907) & 7.8 & 100 \\
Step Drop & 0.889 (0.847--0.920) & 6.9 & 99  \\
\bottomrule
\end{tabular}
\end{table}

\subsection{Self-Selected Locomotor Speed Consistency}

Timing-gate measurements were collected for all cyclic locomotor trials (walking, jogging, and running). Mean speeds were $1.32 \pm 0.18$ m/s for walking, $2.47 \pm 0.34$ m/s for jogging, and $3.71 \pm 0.52$ m/s for running. Within-participant coefficients of variation ranged from 4.1--9.3\%, indicating consistent self-selected pacing across repeated trials.

\paragraph{Speed data availability note.}
Timing-gate measurements were collected for all locomotor trials; however, \textbf{183 trials have valid speed measurements in the released dataset} after excluding trials where the corresponding sensor recordings were unavailable or removed during quality-control curation. Speed measurements are therefore provided only for trials where the associated multi-sensor recording is retained. Timing gates were not used for the \textbf{impact tasks (heel drops and step drops)}. Users may identify trials with valid speed data using the \textbf{speed\_available flag in trial\_manifest.csv}.

\subsection{Force Plate Data}

Vertical GRF signals were visually inspected for completeness and assessed for physiological plausibility. All recordings displayed clear stance phases and expected loading patterns consistent with published biomechanics literature \cite{winter2009,cavanagh1980}. Signal dropout or artefacts led to trial exclusion as documented in the quality control procedures.

Peak vGRF values across activities aligned with published biomechanics literature. Walking trials showed peak forces of 1.1--1.2$\times$ body weight, consistent with normative data \cite{winter2009}. Running exhibited peaks of 2.0--2.5$\times$ body weight, matching values reported by Cavanagh \& Lafortune \cite{cavanagh1980}. Drop landing activities produced peak forces of 2.5--4.0$\times$ body weight, within expected ranges for impact activities.

\subsection{Wearable Sensor Data}

IMU recordings were evaluated for completeness, sampling rate stability and noise characteristics. The applied low-pass filters removed high-frequency noise without distorting locomotor signals. Device timestamps were checked for continuity to ensure no missing samples. Acceleration magnitude computation provided orientation-invariant measures suitable for cross-sensor comparison.

\subsection{Signal Quality}

IMU signal quality was assessed through multiple metrics to ensure data integrity. Sampling rate stability was verified by examining timestamp intervals across all recordings. The Apple Watch sensors maintained consistent sampling with mean interval $10.1 \pm 0.3$ ms (corresponding to approximately 99 Hz), with no systematic drift or gaps exceeding 50 ms detected across the dataset.

Noise characteristics were evaluated by examining signal stability during standing calibration periods at the beginning of each session. During 5-second standing periods, acceleration magnitude showed root-mean-square deviations of $0.008 \pm 0.002$ g for wrist sensors and $0.006 \pm 0.001$ g for waist sensors, indicating acceptable baseline noise levels for movement detection. Gyroscope signals exhibited baseline variability of $0.03 \pm 0.01$ rad/s during standing, confirming stable zero-reference conditions.

The applied low-pass filtering effectively attenuated high-frequency noise while preserving locomotor signal characteristics. Visual inspection of filtered signals confirmed removal of sampling artifacts and electromagnetic interference without introducing phase distortion or signal damping in the frequency range of interest (0--10 Hz for locomotor activities). Force plate signals similarly showed clean vertical GRF profiles with minimal baseline noise ($<$ 5 N during unloaded periods).

Temporal alignment between sensors was achieved using a hybrid alignment procedure (event-based marker detection where available, and stance-feature alignment with cross-correlation validation). Timing uncertainty at the IMU sampling resolution was characterised via Monte Carlo perturbations of $\pm$10 ms ($\pm$1 sample at $\sim$100 Hz), showing minimal impact on correlation-based validation metrics for comparisons with $|r| \geq 0.2$. This analysis quantifies metric sensitivity to bounded lag perturbations rather than establishing absolute timing accuracy, which would require independent ground-truth event timing.

\subsection{Three-Phase Cross-Sensor Validation}

Dataset quality was characterised using a three-phase cross-sensor plausibility and consistency framework designed to summarise relationships between wearable IMU signals and force-plate measurements. Phase 1 (Waist$\to$Force Plate) assesses baseline plausibility of sensor--force coupling using a proximal placement near the trunk; Phase 2 (Wrist$\to$Waist) evaluates agreement between distal and proximal body locations; and Phase 3 (Wrist$\to$Force Plate) investigates the feasibility of wrist-only deployment for vGRF inference. Pearson correlation is used here as an interpretable plausibility and consistency metric for quality-control purposes only; it does not represent validation of GRF estimation accuracy nor assume a purely linear IMU--GRF relationship for model development. Table \ref{tab:dataset_overview} summarises the key characteristics of the dataset.

\begin{table}[H]
\centering
\caption{Dataset overview and key characteristics.}
\label{tab:dataset_overview}
\begin{tabularx}{\textwidth}{lX}
\toprule
\textbf{Characteristic} & \textbf{Description} \\
\midrule
Subject area & Data Science, Biomedical Engineering, Sports Science \\
More specific subject area & Ground reaction force estimation using consumer wearable sensors \\
Type of data & Time series (force plate, IMU accelerometry, gyroscope), metadata (JSON), trial identifiers (UUID) \\
How data were acquired & AMTI OR6-7 force plate (1000 Hz); Apple Watch Series 5+ accelerometers/gyroscopes ($\approx$100 Hz) \\
Data format & Raw and processed (CSV files); Python 3.9+ analysis scripts \\
Parameters for data collection & 10 healthy adults (26--41 years, 6 male/4 female), 5 activities, 492 validated trials \\
Description of data collection & Participants performed walking, jogging, running, heel drops, and step drops with dual Apple Watch placement (wrist, waist). Temporal alignment using a hybrid event-based approach (heel-tap/stance-feature detection) with cross-correlation validation; robustness to single-sample timing jitter ($\pm$10ms at 100 Hz) quantified via Monte Carlo sensitivity analysis. Butterworth filtering: 20 Hz (force), 10 Hz (IMU). Self-selected locomotor speeds measured using timing gates (speed = distance/time, m/s). \\
Data source location & University of Hull, Kingston upon Hull, UK (53.7676$^\circ$N, 0.3274$^\circ$W) \\
Data accessibility & Zenodo concept DOI: 10.5281/zenodo.17376716; Version DOI (v1.0): 10.5281/zenodo.17376717; License: CC-BY-4.0 \\
\bottomrule
\end{tabularx}
\end{table}

To systematically evaluate sensor--GRF relationships and cross-sensor information transfer, we implemented a three-phase correlation analysis framework using the triad-complete subset of the released dataset ($n = 395$ trials). Pearson $r$ is used here as an interpretable plausibility and consistency metric for quality-control purposes only; it does not constitute a validation of sensor-based GRF estimation accuracy and should not be interpreted as a measure of concurrent validity between Apple Watch signals and force plate GRF. Non-linear models (e.g., deep learning) may identify structure that linear correlation does not capture, and low $r$ values do not imply the dataset is unsuitable for model development. We computed Pearson correlations between acceleration magnitude and vGRF over stance for 395 triad-complete trials across three sensor pairings:

\begin{itemize}
  \item \textbf{Phase 1 (Waist$\to$Force Plate):} Mean $r = 0.550$ (range: $-0.21$ to $0.93$).
  \item \textbf{Phase 2 (Wrist$\to$Waist):} Mean $r = 0.486$ (range: $-0.46$ to $0.94$).
  \item \textbf{Phase 3 (Wrist$\to$Force Plate):} Mean $r = 0.486$ (range: $-0.28$ to $0.88$).
\end{itemize}

Activity-specific correlation patterns are shown in Table \ref{tab:cross_sensor}. Locomotor tasks generally exhibited stronger Phase 1 and Phase 3 correlations than impact tasks, consistent with cyclic versus non-cyclic movement patterns.

\begin{table}[H]
\centering
\caption{Activity-specific cross-sensor validation performance for Phase 1 and Phase 3 (395 triad-complete trials from primary manifest).}
\label{tab:cross_sensor}
\begin{tabular}{lccc}
\toprule
\textbf{Activity} & \textbf{N} & \textbf{Phase 1 Mean $r$} & \textbf{Phase 3 Mean $r$} \\
\midrule
Walking   & 85  & 0.543 & 0.430 \\
Jogging   & 71  & 0.608 & 0.582 \\
Running   & 79  & 0.624 & 0.558 \\
Heel Drop & 79  & 0.470 & 0.294 \\
Step Drop & 81  & 0.442 & 0.548 \\
Overall   & 395 & 0.550 & 0.486 \\
\bottomrule
\end{tabular}
\end{table}

\paragraph{Additional timing-sensitive QC metrics (alignment lag and peak-time agreement).}
To complement correlation (which is scale-insensitive and may be misinterpreted as a validity metric), we report two additional timing-sensitive QC descriptors per trial in the alignment logs: (i) the estimated lag (in ms) selected by the alignment procedure, reflecting the inferred temporal offset between force plate and IMU signals; and (ii) the peak-time difference (in ms) between the dominant stance-phase peak in Force\_Z and the corresponding peak in the IMU proxy signal after alignment. These metrics characterise alignment behaviour and timing consistency without implying gold-standard hardware synchronisation. Users can inspect the distribution of lags and peak-time differences by activity and participant to identify atypical trials prior to model development.

\subsection{Time-Alignment Sensitivity Analysis}

This sensitivity analysis was performed on the 598 aligned outputs (post-alignment, pre-quality-control curation), not on the 492 released validated trials, to characterise alignment robustness independently of subsequent exclusions.

To assess robustness of correlation-based validation metrics to residual timing uncertainty at the sampling resolution (100 Hz, 10 ms per sample), we performed Monte Carlo simulations ($n = 100$ perturbations per comparison) applying random lag perturbations of $\pm$10ms ($\pm$1 sample). This analysis quantifies sensitivity of the reported correlations to bounded timing variations rather than establishing absolute timing accuracy, which would require independent ground-truth event timing.

Across 598 trials (1,794 trial-phase comparisons), the overall mean of the per-comparison mean absolute change in correlation was 0.0111, with 95th percentile = 0.0333 and maximum = 0.1624 (Table \ref{tab:mc_sensitivity}). Restricting to comparisons with $|r| \geq 0.2$ ($n=957$, 53.3\%) yielded mean $|\Delta r| = 0.0103$, 95th percentile = 0.0306, maximum = 0.1491 (Figure 4).

For comparisons with $|r| \geq 0.2$, only 1.7\% of Phase 1 comparisons, 11.2\% of Phase 2 comparisons, and 5.0\% of Phase 3 comparisons exceeded the 0.03 sensitivity threshold; correspondingly, 0.0\%, 2.5\%, and 1.0\% exceeded the 0.05 threshold (Figure 6). The moderately elevated Phase 2 sensitivity reflects biomechanical independence between wrist and waist during dynamic activities, particularly jogging and running. Activity-specific sensitivity patterns are illustrated in Figure 7. Jogging showed the highest Phase 2 (wrist$\to$waist) sensitivity (mean $|\Delta r| = 0.017$), consistent with greater biomechanical independence between distal and proximal body locations during higher-intensity cyclic locomotion. Drop landing activities exhibited moderate sensitivity, particularly in the wrist$\to$force-plate comparison, while walking and running demonstrated consistently low sensitivity across phases. Overall, these patterns confirm that correlation-based validation metrics remain robust to single-sample timing perturbations even during activities with more complex inter-segmental coordination.

Empirical cumulative distribution analysis confirmed that $> 95\%$ of comparisons across all phases achieved mean $|\Delta r| < 0.05$ (Figure 5). Among the 133 comparisons (7.4\% of total) with elevated sensitivity ($|\Delta r| > 0.05$) in the full dataset, 53.4\% (71/133) had $|\text{baseline }r| < 0.2$, indicating that large fractional changes concentrate where correlations are weak. Overall, single-sample timing perturbations ($\pm$10ms at 100 Hz) produce minimal changes in correlation-based validation metrics for biomechanically meaningful relationships ($|r| \geq 0.2$), supporting alignment robustness for correlation-based analyses. We verified that manuscript counts (678$\to$598 aligned outputs; 492 validated; 395 triad-complete) match the released dataset manifests.

\begin{table}[H]
\centering
\caption{Monte Carlo alignment sensitivity summary (598 trials; 1,794 comparisons; $\pm$10ms perturbations). P95 denotes the 95th percentile across comparisons of per-comparison mean $|\Delta r|$. Phase rows summarise mean $|\Delta r|$ over comparisons within each phase (598 comparisons per phase).}
\label{tab:mc_sensitivity}
\begin{tabular}{lcccc}
\toprule
\textbf{Subset} & \textbf{N} & \textbf{Mean $|\Delta r|$} & \textbf{P95 $|\Delta r|$} & \textbf{Max $|\Delta r|$} \\
\midrule
All comparisons             & 1,794 & 0.0111 & 0.0333 & 0.1624 \\
$|r| \geq 0.2$ comparisons & 957   & 0.0103 & 0.0306 & 0.1491 \\
\midrule
\multicolumn{5}{l}{\textit{By phase (all 598 trials; mean over comparisons within phase):}} \\
Phase 1 (Waist$\to$FP)    & 598 & 0.0079 & --- & 0.1212 \\
Phase 2 (Wrist$\to$Waist) & 598 & 0.0144 & --- & 0.1624 \\
Phase 3 (Wrist$\to$FP)    & 598 & 0.0111 & --- & 0.1491 \\
\bottomrule
\end{tabular}
\end{table}

\begin{figure}[H]
  \centering
  \includegraphics[width=\textwidth]{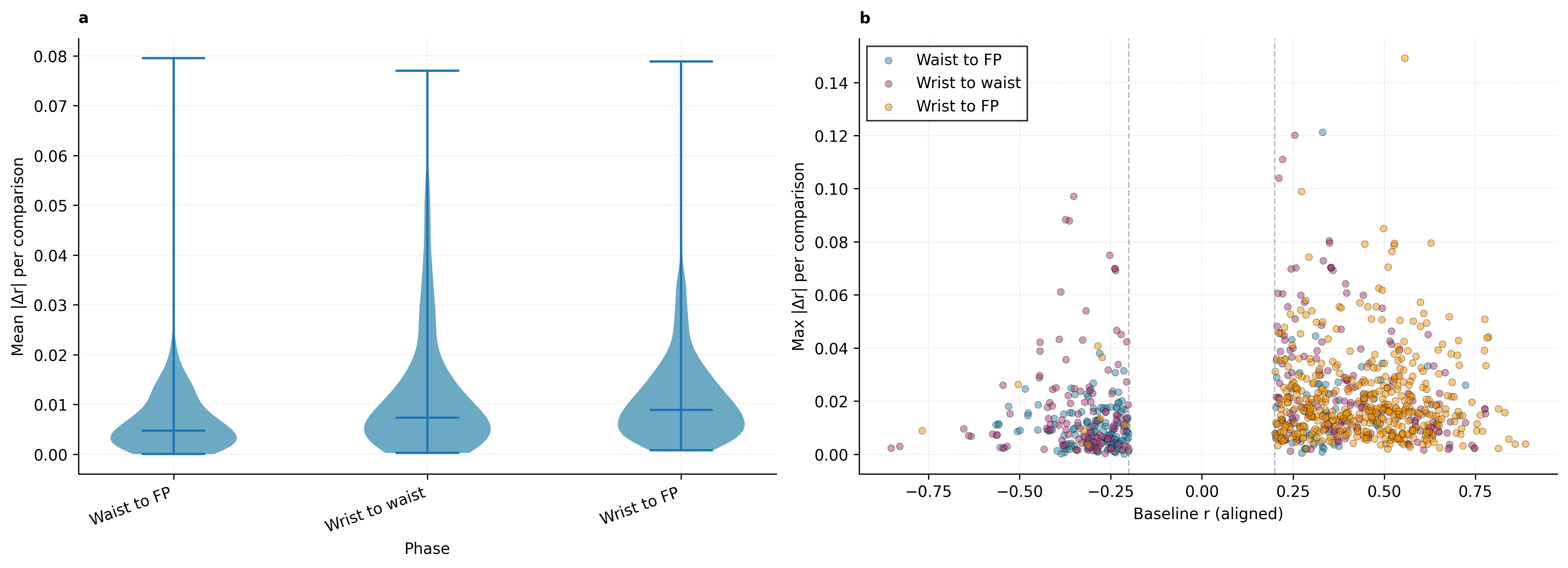}
  \caption{Time-alignment sensitivity overview ($|\text{baseline }r| \geq 0.2$). (A) Distribution of per-comparison mean $|\Delta r|$ by validation phase. Violin plots show kernel density estimates; box plots show median, interquartile range, and outliers. (B) Relationship between baseline correlation and sensitivity. Vertical dashed lines mark the $|r| = 0.2$ threshold; the horizontal dotted line indicates a sensitivity threshold of 0.05.}
  \label{fig:fig4}
\end{figure}

\begin{figure}[H]
  \centering
  \includegraphics[width=0.6\textwidth]{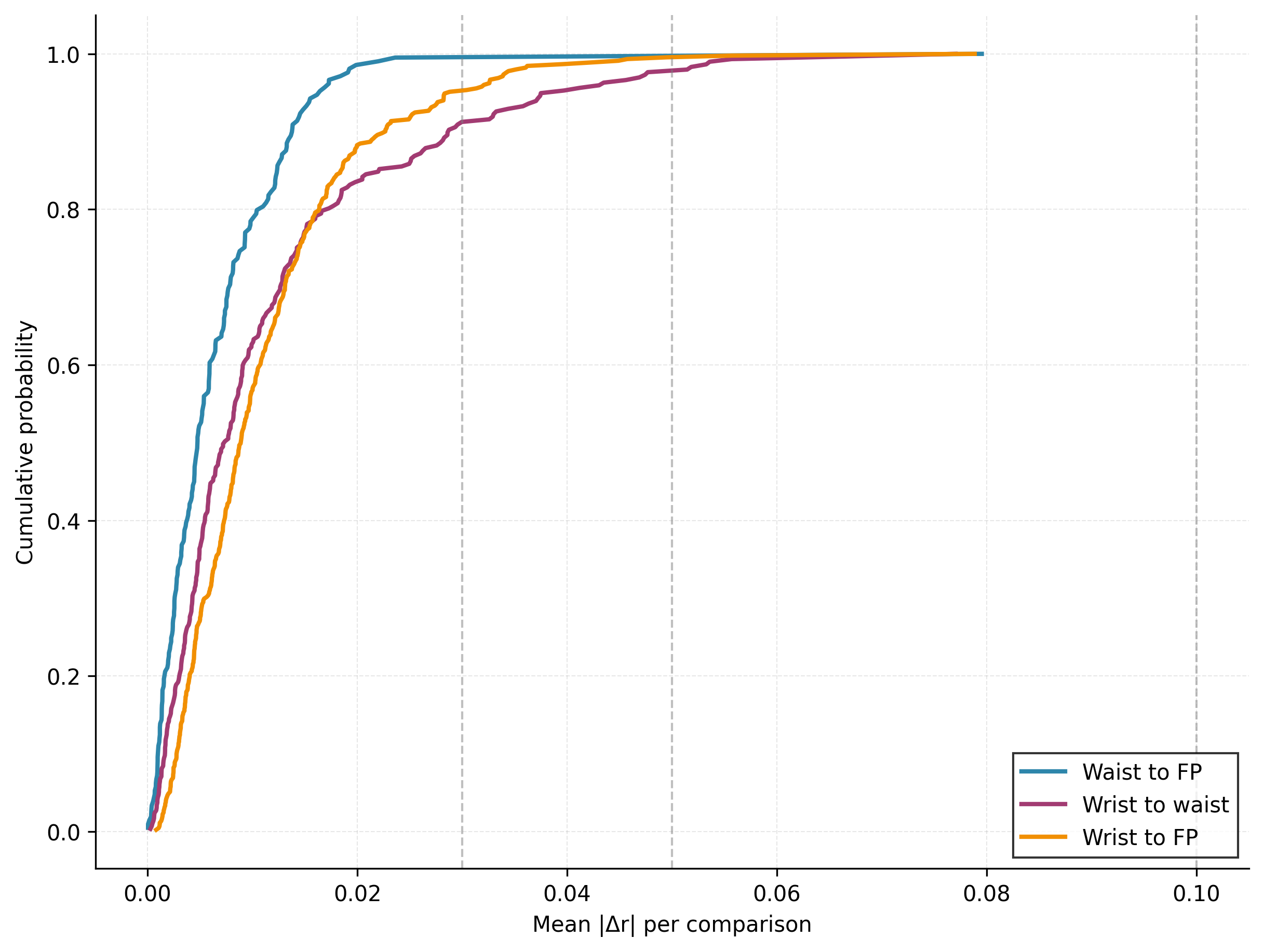}
  \caption{Empirical cumulative distribution of sensitivity by phase ($|\text{baseline }r| \geq 0.2$). Vertical dashed lines mark thresholds of 0.03 and 0.05.}
  \label{fig:fig5}
\end{figure}

\begin{figure}[H]
  \centering
  \includegraphics[width=0.6\textwidth]{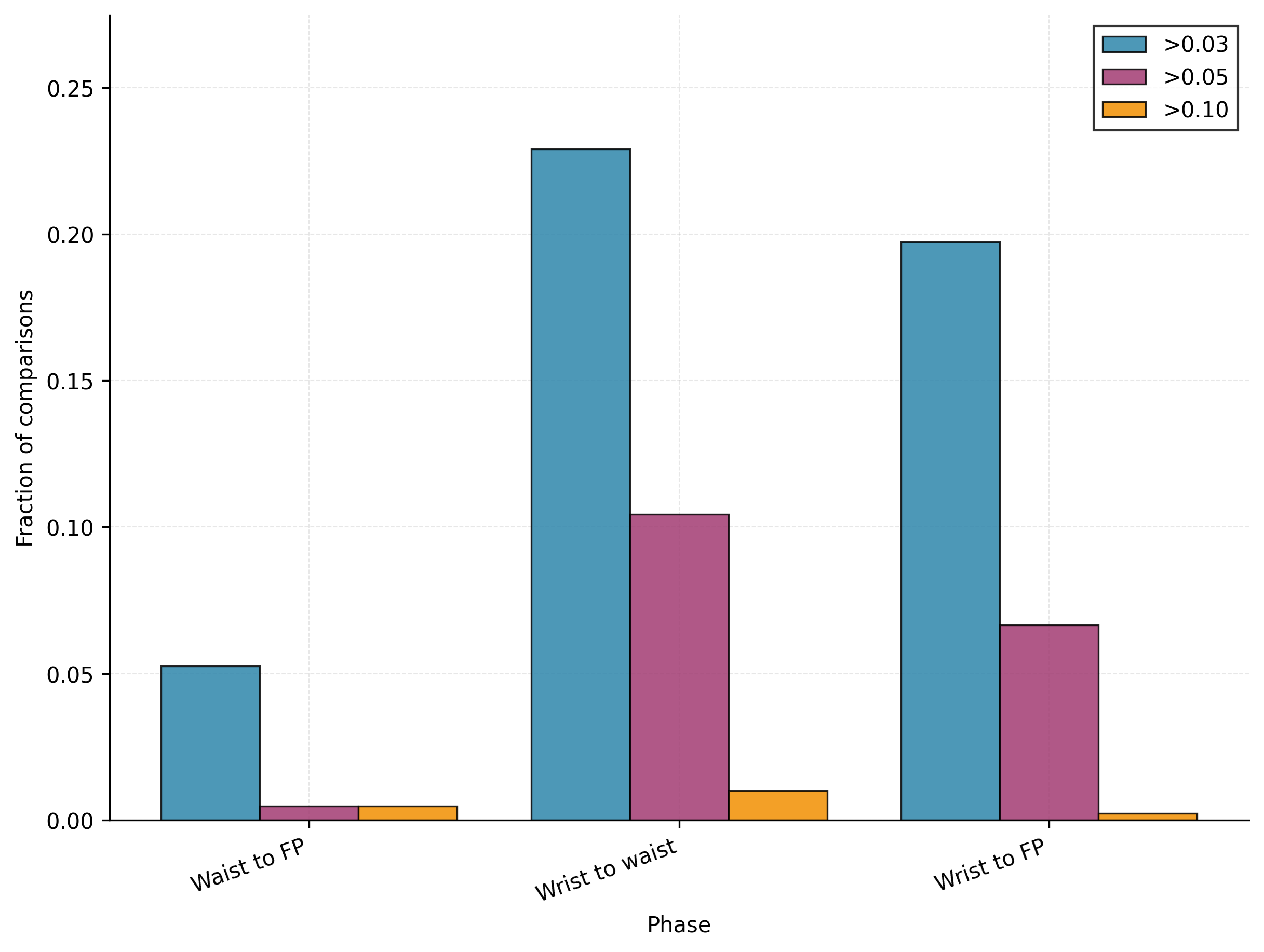}
  \caption{Threshold exceedance rates by phase ($|\text{baseline }r| \geq 0.2$). Bars show the fraction of comparisons exceeding sensitivity thresholds of 0.03, 0.05, and 0.10.}
  \label{fig:fig6}
\end{figure}

\begin{figure}[H]
  \centering
  \includegraphics[width=0.65\textwidth]{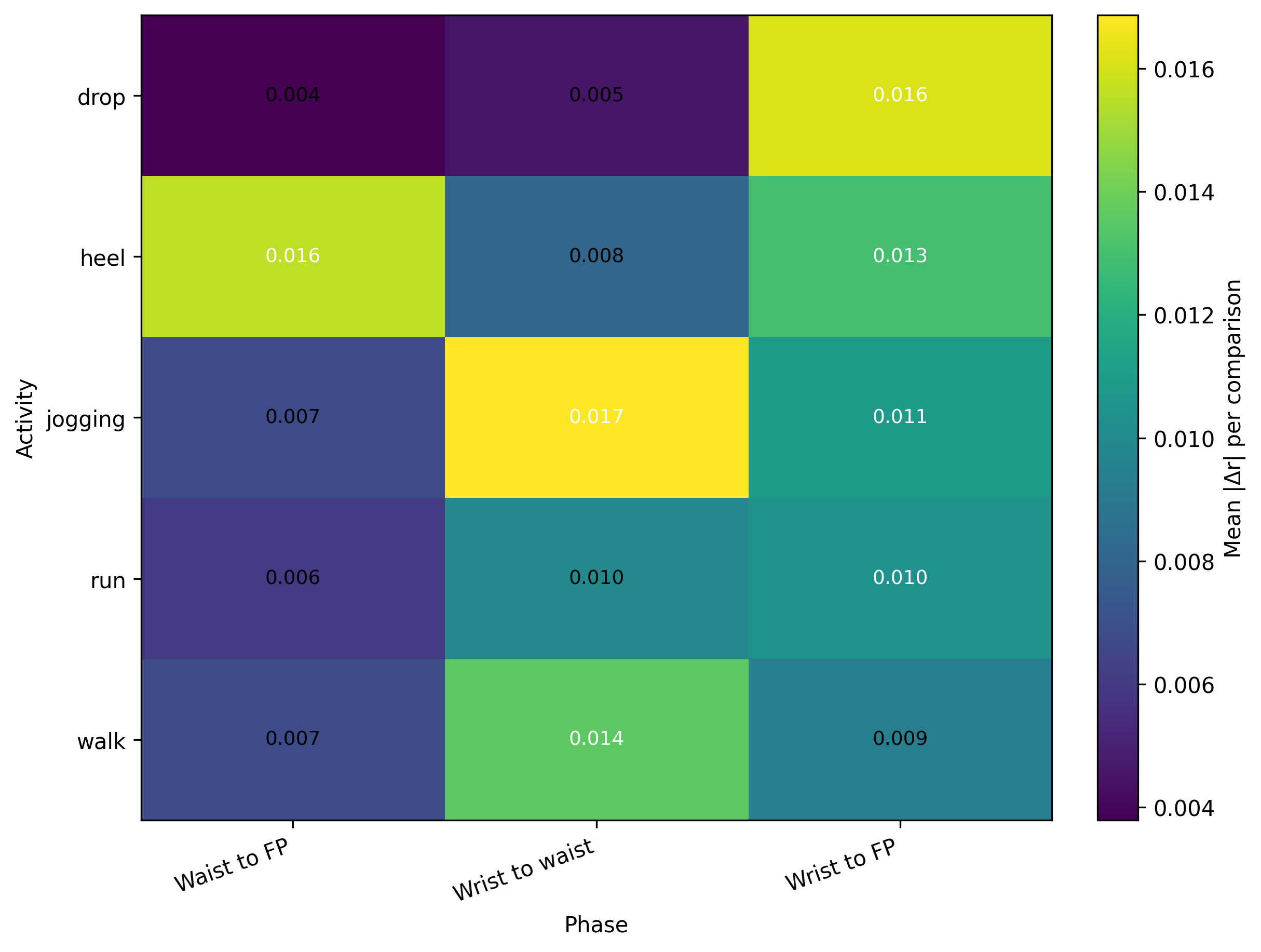}
  \caption{Mean sensitivity by activity and phase ($|\text{baseline }r| \geq 0.2$). Heatmap shows mean $|\Delta r|$ for each activity-phase combination.}
  \label{fig:fig7}
\end{figure}

\subsection{Data Completeness and Missing Modalities}

Sensor availability across the 492 validated trials is high: wrist data for 461 trials (93.7\%), waist data for 451 trials (91.7\%), and force plate data for 449 trials (91.3\%). The triad-complete subset (395 trials, 80.3\%) represents trials where all three modalities are present.

Missing data patterns are not uniformly distributed. Systematic gaps are observed for specific participant-activity combinations: P1 jogging (0/10 triad-complete), P5 heel drops (0/11 triad-complete) and step drops (0/11 triad-complete), and P8 jogging (0/12 triad-complete). These patterns likely reflect technical issues during collection (battery depletion, connectivity interruptions, file system errors) rather than random data loss. Users should account for these patterns when designing subgroup analyses.

% ── Usage Notes ───────────────────────────────────────────────────────────────
\section{Usage Notes}

This dataset supports reproducible research in wearable biomechanics, model development, and sensor placement investigations. We provide recommendations for effective use:

\paragraph{Trial selection.}
The primary manifest (trial\_manifest.csv) contains 492 validated trials; the triad-complete subset (395 trials) is recommended for analyses requiring all three modalities.

\paragraph{Sensor orientation.}
Signals are reported in device coordinates. Acceleration magnitude is provided as an orientation-agnostic feature; users requiring anatomical frames should apply their own transformations.

\paragraph{Time-alignment considerations.}
Aligned signals are provided. Monte Carlo analysis indicates that single-sample timing perturbations ($\pm$10ms at 100 Hz) produce small changes in correlation-based metrics for meaningful baseline correlations ($|r| \geq 0.2$).

\paragraph{Machine learning considerations.}
We recommend participant-stratified evaluation (e.g., Leave-One-Participant-Out or GroupKFold) to assess generalisation and avoid data leakage; random row-wise splits are inappropriate because rows within a trial are temporally autocorrelated and share participant identity. Users should account for missing-modality patterns when defining splits and verify that the missingness mechanism does not differ systematically between train and test partitions. Given the modest sample size (10 participants), both per-participant and pooled performance metrics should be reported, alongside confidence intervals across folds. Drop tasks should be treated separately from cyclic locomotor tasks, with particular attention to saturation flags; unfiltered raw signals are preferred for high-frequency impact analyses. The recommended checklist below summarises key evaluation practices.

\vspace{0.5em}
\noindent\fbox{%
\begin{minipage}{0.95\textwidth}
\textbf{Recommended ML Usage Checklist}
\begin{itemize}[leftmargin=1.5em]
  \item[$\checkmark$] Use participant-level data splits (Leave-One-Participant-Out or GroupKFold). Never use random row-wise splits --- rows within a trial are temporally autocorrelated and share participant identity, causing severe data leakage.
  \item[$\checkmark$] Report per-participant performance metrics (mean $\pm$ SD and confidence intervals across folds), not only pooled metrics. Given $n = 10$ participants, pooled metrics alone may mask high inter-participant variability.
  \item[$\checkmark$] Select the appropriate manifest before splitting use trial\_manifest.csv (triad-complete, $n = 395$) for analyses requiring all three modalities; use the full 492-trial manifest for single-sensor analyses. Verify missing-modality patterns do not differ between train and test sets.
  \item[$\checkmark$] Treat impact tasks (heel drops, step drops) separately from cyclic locomotor tasks. Check saturation flags before including drop trials; prefer unfiltered raw signals for high-frequency impact analyses; avoid aggressive low-pass filtering that attenuates impact transients.
  \item[$\checkmark$] Document any imputation or exclusion strategy for missing modalities and verify that the missingness mechanism is not confounded with participant or activity in your splits.
\end{itemize}
\end{minipage}
}
\vspace{0.5em}

The dataset supports multiple research applications including development of machine learning models for force prediction, investigation of sensor placement effects, benchmarking of signal processing algorithms, activity recognition studies, and methodological research on multi-sensor time-alignment protocols. Example analysis scripts demonstrating data loading, preprocessing, and basic feature extraction are included in the repository to facilitate adoption and reproducible research.

For machine-learning applications, we recommend participant-stratified cross-validation to avoid information leakage between training and test sets. Given the modest sample size (10 participants), users should report both per-participant and pooled performance metrics and consider bootstrapped confidence intervals when benchmarking models to ensure robust and transparent evaluation.

\paragraph{Self-selected speed.}
Timing-gate measurements were collected for all locomotor trials (walking, jogging, and running), yielding 303 speed records in total. These measurements are provided in a separate Excel file and are linked to trials using participant code, activity, and trial number. Of these, 183 trials correspond to locomotor trials retained in the final validated sensor dataset after QC curation, enabling direct comparison between timing-gate speed and the associated wearable and force-plate recordings. Timing gates were not used for the impact tasks (heel drops and step drops).

\paragraph{Filtering and signal processing.}
Filtered signals are provided for convenience (20 Hz force, 10 Hz IMU), but users should access raw signals for analyses requiring higher-frequency content.

\paragraph{Accelerometer saturation.}
Apple Watch accelerometers have a nominal measurement range of $\pm$8.0g. High-impact activities may approach this limit, causing saturation (clipping) that can affect peak acceleration estimates and downstream GRF inference. We performed a systematic saturation analysis across all trials using automated detection of: (i) near-limit flat-top patterns ($\geq$3 consecutive samples at $>$95\% of the nominal range, i.e., $\geq$7.6g), (ii) hard clipping ($\geq$2 consecutive samples at $>$99\% of the nominal range, i.e., $\geq$7.92g), or (iii) $>$0.1\% of samples within 95\% of the nominal range. Overall, suspected saturation was observed in 3.5\% of wrist trials (16/461) and 14.6\% of waist trials (66/451). Notably, waist step drops were frequently affected (48.9\%, 46/94), whereas wrist running showed 12.9\% (12/93) and wrist step drops 4.2\% (4/96). Heel drops exhibited 7.3\% suspected saturation at the waist (7/96) and 0\% at the wrist, while walking and jogging showed minimal saturation. Accounting for trial length, the weighted proportion of near-limit samples was 0.014\% (wrist) and 0.030\% (waist), indicating brief transient excursions rather than sustained clipping. Maximum observed acceleration magnitudes in the analysed files were 10.2g (wrist) and 14.7g (waist); occasional peaks exceeding 8g are most likely explained by the fact that the resultant vector magnitude (Euclidean norm across three axes) can exceed the per-axis ceiling even when no individual axis is saturated --- for example, simultaneous near-limit accelerations on two or three axes produce a resultant up to $\sqrt{3} \times 8\,\mathrm{g} \approx 13.9\,\mathrm{g}$. Additionally, Apple's Core Motion framework may apply internal scaling or sensor fusion corrections that are not fully documented in the public API. Because these sources cannot be definitively resolved from the available device APIs, absolute acceleration magnitudes should not be treated as fully calibrated or directly comparable across trials without appropriate normalisation; users are advised to rely on the provided per-trial saturation flags and to use acceleration magnitude primarily as a relative, within-trial feature rather than an absolute cross-trial measurement. Accordingly, we report saturation relative to proximity to the nominal $\pm$8g limits rather than absolute maxima. Users analysing impact tasks should consider saturation flags and robust feature extraction/estimation methods; unfiltered raw signals are provided to enable custom detection and mitigation.

\paragraph{Limitations.}
This dataset captures controlled laboratory movements and may not represent free-living conditions. The sample size ($n=10$) limits population-level inference but is comparable to similar biomechanics datasets (Table \ref{tab:comparison}). Although the force-plate exports include the full AMTI channels (three-dimensional forces, moments, and centre-of-pressure coordinates), the present dataset curation and validation focus primarily on the vertical ground reaction force component (vGRF; Force\_Z). Users requiring additional channels may extract them directly from the raw force-plate files. The provision of full force-plate channels allows future studies to investigate mediolateral or anteroposterior force components if required.

The current dataset is limited to healthy adults aged 26--41 years recruited from a single laboratory site. Generalisability to elderly populations, clinical cohorts, or individuals with diverse anthropometric characteristics requires further validation. Data collection was performed under controlled laboratory conditions with standardised surfaces and structured activity protocols. Real-world deployment scenarios will encounter greater variability, including diverse footwear types, varied terrain surfaces, and unpredictable movement transitions. Footwear was not standardised; however, shoe type and shoe size were recorded for all participants and are provided in the accompanying metadata, allowing users to stratify analyses if required. Participants wore their own athletic footwear and completed the protocol on a standard laboratory floor surface. The five locomotor activities represent common movement patterns but do not capture sport-specific tasks or the full range of activities of daily living. Finally, the dataset includes devices from a single wearable manufacturer (Apple Watch), additional cross-device validation will be required to generalise findings to other wearable sensor platforms.

% ── Code Availability ─────────────────────────────────────────────────────────
\section{Code Availability}

All preprocessing and analysis scripts used to generate the released dataset, perform temporal alignment, run sensitivity analyses, and reproduce figures are archived on Zenodo and mirrored on GitHub: \\\url{https://github.com/ParvinGhaffarzadeh/Apple-Watches-and-Force-Plate-dataset}. The repository includes scripts for (i) temporal alignment, (ii) Monte Carlo time-alignment sensitivity analysis, (iii) manifest reconstruction and UUID-based trial matching, (iv) filtering/resampling and feature extraction (e.g., acceleration magnitude), (v) correlation-based quality control, and (vi) figure reproduction. Example Jupyter notebooks demonstrating end-to-end workflows are also provided. The code is implemented in Python ($\geq$3.9) and uses standard scientific libraries (NumPy, SciPy, Pandas, Matplotlib).

\paragraph{Reproducibility notes:}
Alignment was executed on 678 matched force+waist+wrist recordings, yielding 598 aligned outputs. The released dataset contains 492 validated trials (395 triad-complete) after additional curation. Time-alignment sensitivity analysis was performed on aligned outputs; three-phase validation uses the 395 triad-complete released trials.

% ── Data Availability ─────────────────────────────────────────────────────────
\section{Data Availability}

The dataset is publicly available on Zenodo: \url{https://doi.org/10.5281/zenodo.17376717} (concept DOI: \url{https://doi.org/10.5281/zenodo.17376716}) under a Creative Commons Attribution 4.0 International (CC BY 4.0) license.

\section*{Acknowledgments}

We thank all study participants for their time and contribution to this research. We acknowledge University of Hull research facilities and Prof.\ Grant Abt for equipment access and technical support.

\section*{Author Contributions}

P.G.\ designed the study, collected experimental data, performed data processing and statistical analyses, created visualisations, and wrote the manuscript. D.C.\ designed the study, contributed to study conceptualisation, provided supervision, and edited the manuscript. K.A.\ designed the study and edited the manuscript. A.D.\ designed the study, provided resources and edited the manuscript. Y.P.\ designed the study, provided supervision, and edited the manuscript.

\section*{Competing Interests}

The authors declare no competing interests.

\section*{Funding}

This research received no specific grant funding from agencies in the public, commercial, or not-for-profit sectors.

% ── References ────────────────────────────────────────────────────────────────

\end{document}